\documentclass[structabstract]{aa}

\usepackage{url}
\usepackage[breaklinks=true]{hyperref}
\usepackage{twoopt}
\usepackage{natbib}
\bibpunct{(}{)}{;}{a}{}{,} 
\usepackage{ctable}
\usepackage{multirow}
\usepackage{bigstrut}
\usepackage[farskip=0pt]{subfig}
\usepackage[varg]{txfonts}

\def \deg         {$^{\circ}$}
\def \arcmin      {$^\prime$}
\def \arcsec      {$^{\prime\prime}$}
\def \hour        {$^{\mathrm{h}}$}
\def \min         {$^{\mathrm{m}}$}
\def \sec         {$^{\mathrm{s}}$}
\def \mjybeam     {mJy\,beam$^{-1}$}

\def \klambda     {k$\lambda$}

\newcommand{\beam}[2]{{#1}\arcsec $\times$ {#2}\arcsec}

\newcommand{\am}[2]{$#1'\,\hspace{-1.mm}.\hspace{.1mm}#2$}
\newcommand{\Hi}{H\textsc{i}}

\def \vla           {\emph{VLA}}

\def \gmrt          {GMRT}

\def \target       {SNR G351.0-5.4}

\begin{document}

\title{Discovery of the supernova remnant G351.0-5.4}
\titlerunning{The supernova remnant G351.0-5.4}

\author{F.~de~Gasperin\inst{1}
\and C. Evoli\inst{2}
\and M.Br\"uggen\inst{1}
\and A. Hektor\inst{3}
\and M. Cardillo\inst{4}
\and P. Thorman\inst{5}
\and W.A. Dawson\inst{5,6}
\and C.B. Morrison\inst{5,7}
}

\authorrunning{F.~de~Gasperin et al.} 
\offprints{F.~de~Gasperin}

\institute{$^1$Universit\"at Hamburg, Hamburger Sternwarte, Gojenbergsweg 112, 21029, Hamburg, Germany \email{fdg@hs.uni-hamburg.de}\\
$^2$Institut f\"ur Theoretische Physik, Universit\"at Hamburg, Luruper Chaussee 149, 22761, Hamburg, Germany\\
$^3$National Institute of Chemical Physics and Biophysics, Ravala 10, Tallinn, Estonia\\
$^4$INAF-Osservatorio di Arcetri, Largo Enrico Fermi 5, 50125, Firenze, Italy\\
$^5$University of California, One Shields Avenue, Davis, CA 95616, USA\\
$^6$Lawrence Livermore National Lab, 7000 East Avenue, Livermore, CA 94550, USA\\
$^7$Argelander Institute for Astronomy, University of Bonn, Auf dem H\"ugel 71, 53121 Bonn, Germany
}

\date{}
\abstract
{While searching the NRAO VLA Sky Survey (NVSS) for diffuse radio emission, we have serendipitously discovered extended radio emission close to the Galactic plane. The radio morphology suggests the presence of a previously unknown Galactic supernova remnant. An unclassified $\gamma$-ray source detected by EGRET (3EG J1744-3934) is present in the same location and may stem from the interaction between high-speed particles escaping the remnant and the surrounding interstellar medium.}
{Our aim is to confirm the presence of a previously unknown supernova remnant and to determine a possible association with the $\gamma$-ray emission 3EG J1744-3934.}
{We have conducted optical and radio follow-ups of the target using the Dark Energy Camera (DECam) on the Blanco telescope at Cerro Tololo Inter-American Observatory (CTIO) and the Giant Meterwave Radio Telescope (GMRT). We then combined these data with archival radio and $\gamma$-ray observations.}
{While we detected the extended emission in four different radio bands (325, 1400, 2417, and 4850 MHz), no optical counterpart has been identified. Given its morphology and brightness, it is likely that the radio emission is caused by an old supernova remnant no longer visible in the optical band. Although an unclassified EGRET source is co-located with the supernova remnant, Fermi-LAT data do not show a significant $\gamma$-ray excess that is correlated with the radio emission. However, in the radial distribution of the $\gamma$-ray events, a spatially extended feature is related with SNR at a confidence level $\sim1.5\sigma$.}
{We classify the newly discovered extended emission in the radio band as the old remnant of a previously unknown Galactic supernova: \target.}

\maketitle 

\section{Introduction}
\label{sec:introduction}

Supernova Remnants (SNRs) have a fundamental role inside our Galaxy and in general, inside their host-galaxies. The gravitational collapse of a stellar core causes the ejection of the outer layers of the star at very high velocities ($\approx 10, 000$ km s$^{-1}$). The interaction between the circumstellar medium and the ejected matter produces shock-waves propagating at a non-negligible fraction of the speed of light ($v \approx 0.1-0.2$ c).
In addition, diffusive shock acceleration (DSA) predicts that shock fronts observed in SNRs are efficient accelerators of Galactic Cosmic Rays (GCR). Accelerated protons interact with the protons of the ISM producing a $\pi^0$ meson that then decays into two $\gamma$-ray photons \citep[][and references therein]{Malkov2001}. At the same time, accelerated electrons will follow a power-law momentum distribution and produce radio non-thermal emission from the SNR. The proton-proton interaction time decreases with the target density enhancement. Consequently, $\gamma$-ray emission from SNRs can be more easily seen when the remnant interacts with a target denser than the ISM, such as a molecular cloud, as is the case with S147 \citep{Katsuta2012}, IC443 \citep{Tavani2010, Ackermann2013}, W28 \citep{Giuliani2010} and W44 \citep{Ackermann2013, Cardillo2014}.


It has been predicted \citep{Berkhuijsen1984, Tammann1994} that the total number of Galactic SNRs is between 1000 and $10, 000$. However, only $\sim 274$ SNRs have been identified \citep[][and references therein]{Green2009}. Most of the known SNRs are sources of radio-synchrotron emission and radio observations were routinely used to discover new candidates. Images from radio surveys, such as the NRAO VLA Sky Survey \citep[NVSS;][]{Condon1998}, were successfully used in the past to identify new SNR candidates \cite[see e.g.,][for the case of G353.9-2.0]{Green2001}. The NVSS provides polarized images at 1.4 GHz, covering the sky down to a declination of $-40$\deg{}, and its data were collected with the VLA in D-configuration. Although the survey images are obtained with short snap-shots which may limit the sensitivity on the largest scales, the NVSS is a good starting point to look for extended radio emission, up to scales of $\sim16$\arcmin.

By inspecting NVSS radio maps, we discovered a new candidate SNR which we designated \target{}. Here we present a radio follow-up at 325 MHz using the GMRT and an optical follow-up using DECam on the CTIO 4~m Blanco telescope. Furthermore, \target{} appears to be co-located with the unclassified EGRET $\gamma$-ray source 3EG J1744-3934. In order to test the EGRET detection, we have conducted a search for $\gamma$-ray emission at the location of \target{} with the Large Area Telescope (LAT) on board the Fermi Gamma-Ray Space Telescope.

In Sec.~\ref{sec:observations} we present the GMRT (Sec.~\ref{sec:GMRT}), DECam (Sec.~\ref{sec:DECam}) and Fermi (Sec.~\ref{sec:fermi}) observations and the respective data reductions. The main characteristics of \target{} are presented in Sec.~\ref{sec:target}. Discussion and conclusions are in Sec.~\ref{sec:discussion} and Sec.~\ref{sec:conclusions}, respectively.

\section{Observations and data reduction}
\label{sec:observations}

\subsection{GMRT}
\label{sec:GMRT}

Using the \gmrt{}\footnote{\gmrt{}: \url{http://gmrt.ncra.tifr.res.in}}, we observed the radio emission in the direction of \target{} with a run of 5 hours. During the night of March 18--19, 2013, visibilities were recorded at 325~MHz in dual-polarization (RR and LL). We used an 8~second integration time, and a 33~MHz bandwidth divided into 512 channels. The (primary) flux and bandpass calibrator 3C~286 was observed at the start and end of each run, while the (secondary) gain calibrator PKS 1827-36 was observed every half an hour for three minutes. The total time-on-target was 135 minutes.

The visibility data was processed using a semi-automated CASA\footnote{CASA: \url{http://casa.nrao.edu}}-based data reduction pipeline written in Python. Data were initially flagged through a series of increasingly sensitive runs of the CASA tasks FLAGDATA in the ``rflag'' mode. During these cycles a refined bandpass was obtained. 3C~286 was used as the flux and bandpass calibrator. Flags were then reset and, once corrected for the bandpass, AOflagger \citep{Offringa2012} was used to detect and remove radio frequency interference. At the end of the procedure seven antennas (two of which in the core) were manually flagged together with approximately 45\% of the remaining data because of interference. After an initial phase and amplitude calibration on 3C~286, bad baselines were identified and removed by inspecting the BLCAL calibration tables, and the phase and amplitude solutions were extended to the target field. Several cycles of phase self-calibration, with subsequent clipping of the outliers, were performed and a final cycle of amplitude self-calibration was eventually made. Calibration in time and amplitude in the direction of the four brightest sources in the field and subsequent subtraction \citep[``peeling'', ][]{Noordam2004} was done iteratively after subtracting the best model of the rest of the field. The source to peel is then subtracted and the other sources restored. Location of the peeled sources and flux densities are given in Table~\ref{tab:peel}.

\begin{table}[!t]
\begin{center}
\begin{tabular}{cccc}
Source name & RA (J2000) & Dec (J2000) & $S_{325}$ (Jy) \bigstrut[b] \\
NVSS J174802-385754 & 17\hour48\min01\sec & $-38$\deg57\min51\sec & 0.24  \bigstrut[t]\\
NVSS J174543-392912 & 17\hour45\min44\sec & $-39$\deg29\min12\sec & 0.15 \\
NVSS J174737-394957 & 17\hour47\min38\sec & $-39$\deg49\min59\sec & 0.13 \\
NVSS J174122-392935 & 17\hour41\min22\sec & $-39$\deg29\min35\sec & 0.12 \bigstrut[b]\\
\end{tabular}
\end{center}
\caption{Coordinates and flux densities at 325 MHz of the subtracted point sources}\label{tab:peel}
\end{table}

\begin{figure}
\centering
\includegraphics[width=.5\textwidth]{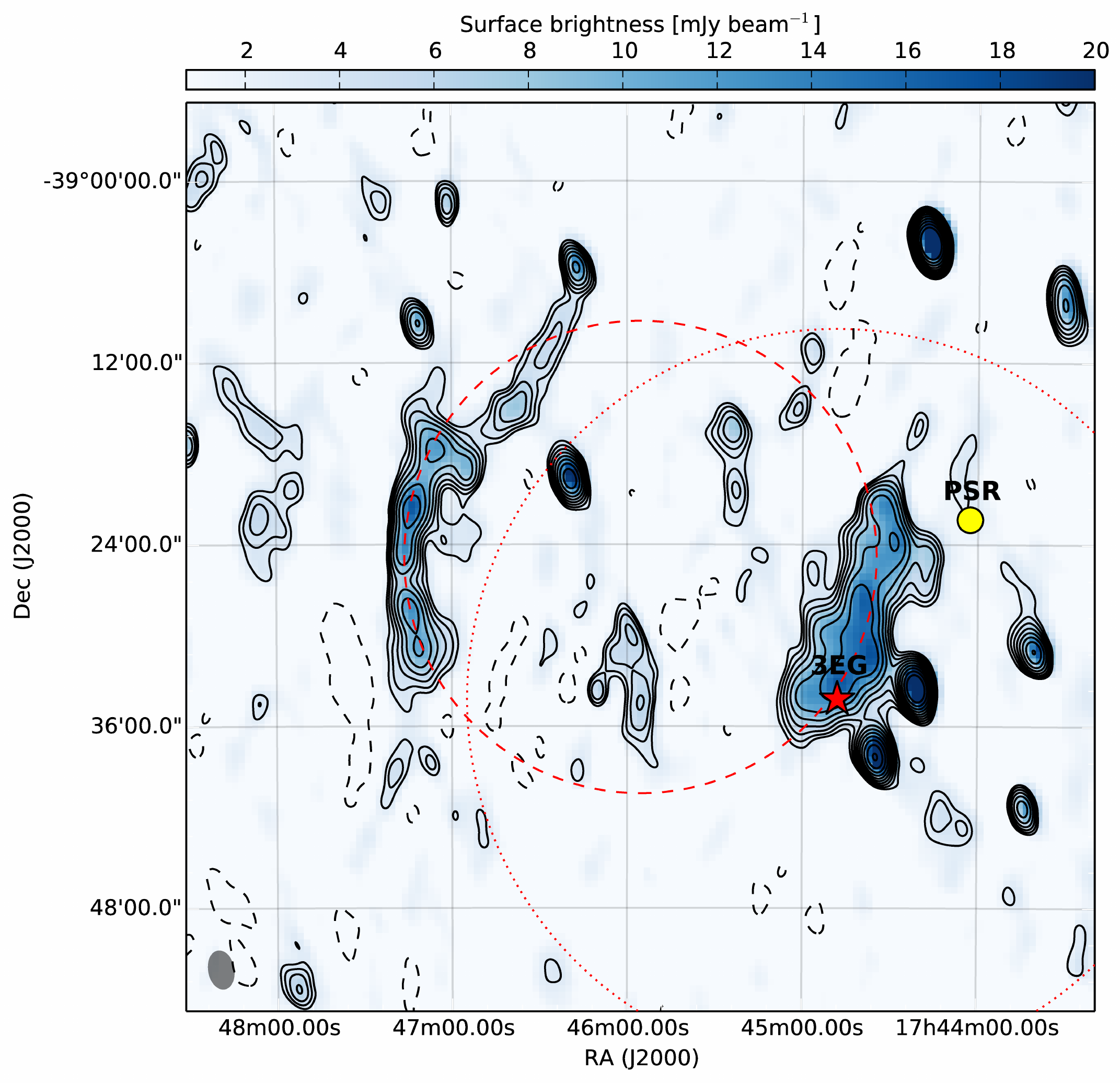}
\caption{GMRT radio maps of \target{} at 325 MHz. Beam size is shown in the bottom left corner, contours are equally spaced between $2\sigma$ and $20\sigma$ in logarithm space with $\sigma = 1.5$~\mjybeam. Dashed contours are at $-2\sigma$. The dashed circle traces the SNR shell location, while the yellow dot shows the positions of the pulsar PSR J1744-3922 and the red star shows the position of the $\gamma$-ray source 3EG J1744-3934. The dotted circle is the $1\sigma$ position error of the $\gamma$-ray source.}\label{fig:GMRT325}
\end{figure}

A final image was made removing baselines longer than 8~\klambda{} and data were tapered applying a circular Gaussian taper with FWHM at 18~\klambda{} to reach a resolution of \beam{141}{89} and enhance the extended emission of the target. The resulting primary-beam corrected image has a background noise of 1.5~\mjybeam{} and is presented in Fig.~\ref{fig:GMRT325}. The flux densities of the east and west part of the shell are listed in Table~\ref{tab:radio}.

\subsection{Archival radio data}
\label{sec:archival_data}

\begin{figure*}
\centering
\subfloat[\vla{} (NVSS) -- 1400 MHz]{\includegraphics[width=.33\textwidth]{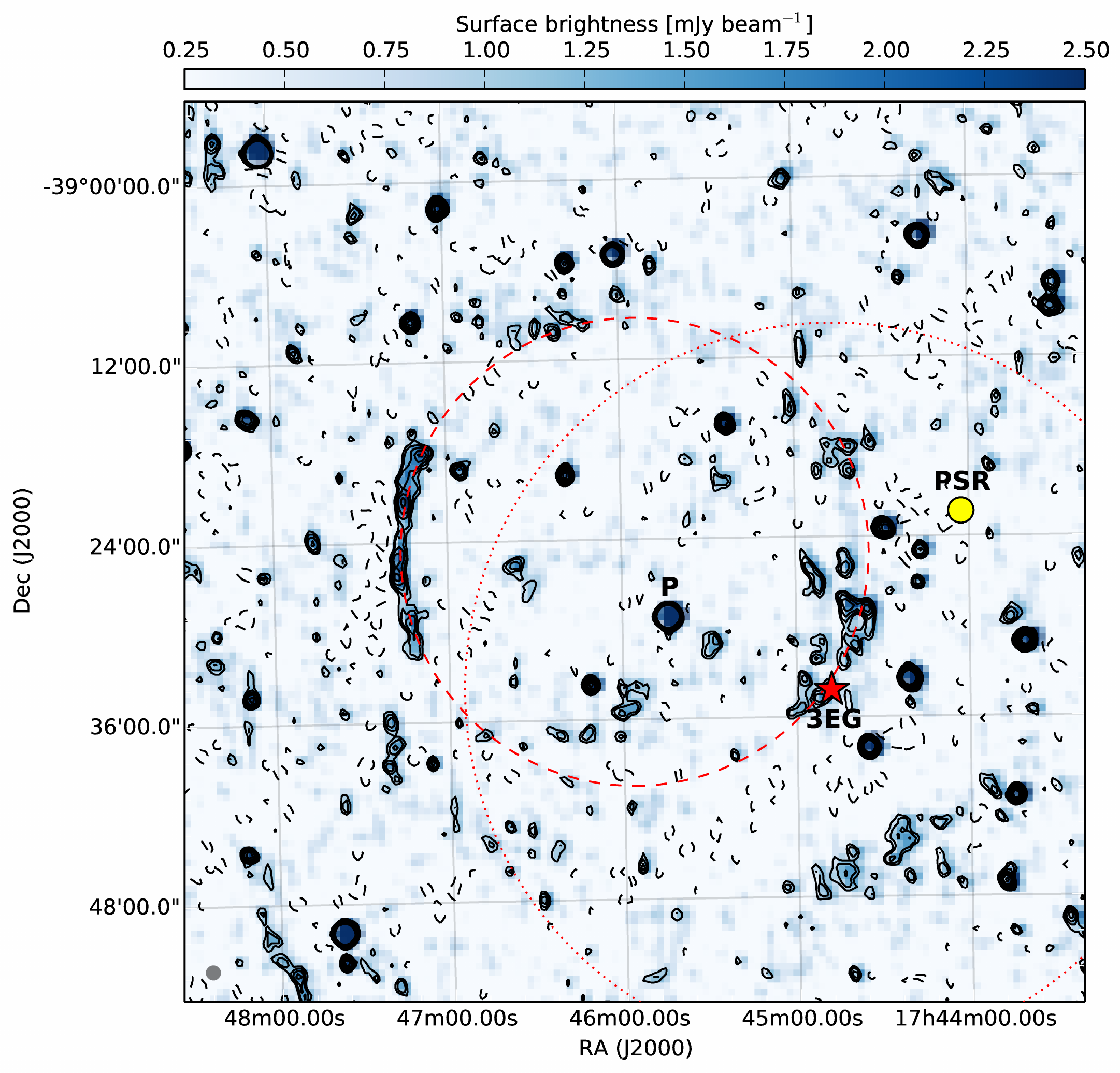}\label{fig:VLA1400}}
\subfloat[Parkes 64m -- 2417 MHz]{\includegraphics[width=.33\textwidth]{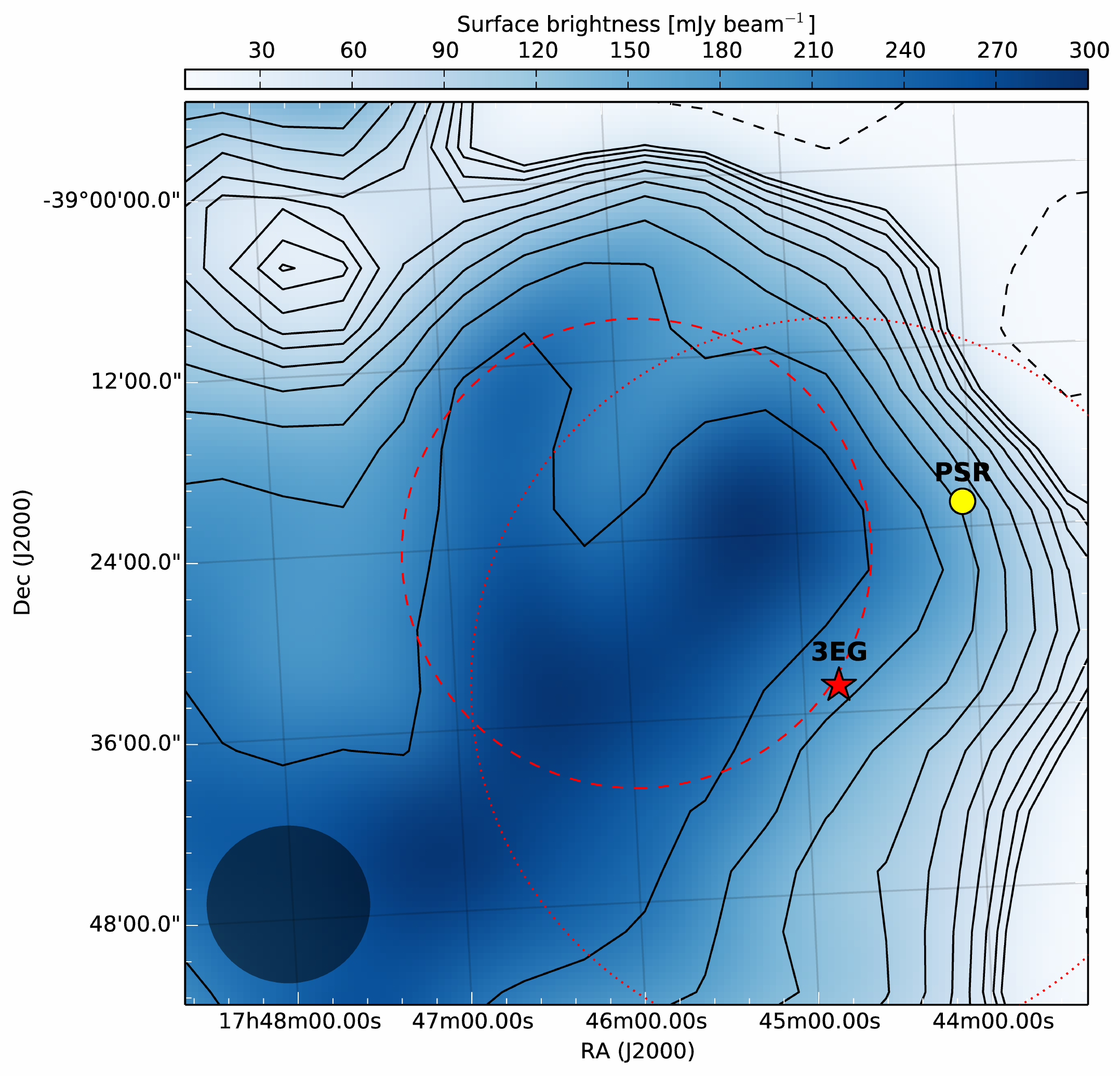}\label{fig:PKS2400}}
\subfloat[Parkes 64m (PMN) -- 4850 MHz]{\includegraphics[width=.33\textwidth]{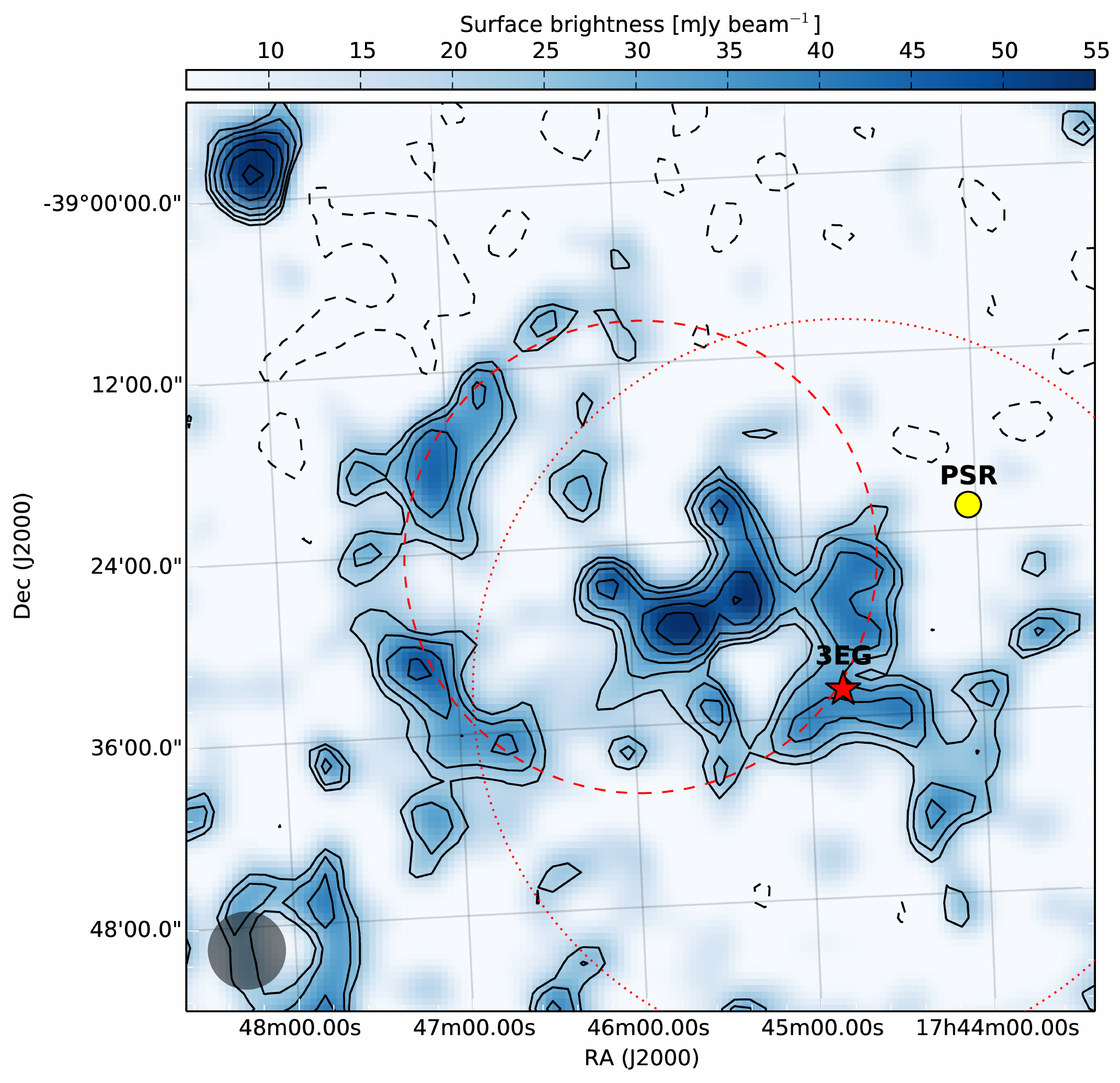}\label{fig:PNM4850}}
\caption{Radio maps of \target{} at 1400 (NVSS), 2417 (Parkes 64m) and 4850 (PMN) MHz. Beam size is shown in the bottom left corner of every panel, contours are equally spaced between $2\sigma$ and $20\sigma$ in logarithm space with $\sigma_{1400} = 0.5$~\mjybeam, $\sigma_{2400} = 11$~\mjybeam, and $\sigma_{4850} = 11$~\mjybeam. Dashed contours are at $-2\sigma$. Markers and lines are as in Fig.~\ref{fig:GMRT325}.}\label{fig:radio}
\end{figure*}

\begin{figure}
\centering
\includegraphics[width=.5\textwidth]{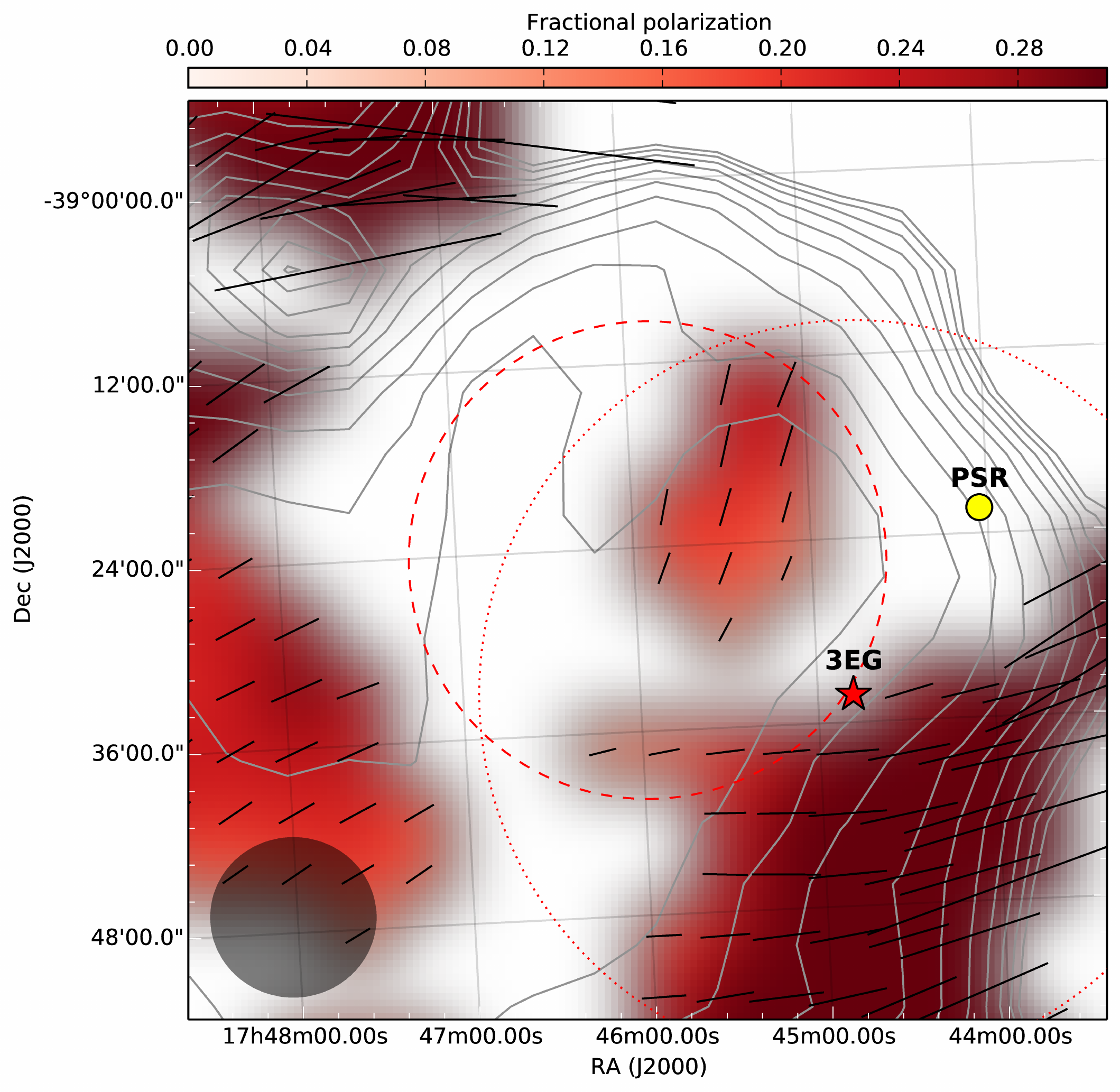}
\caption{Fractional polarization map obtained from the Parkes 2.4 GHz Survey of the southern Galactic Plane. E-vectors (perpendicular to the magnetic field) are also displayed for every pixel of the original maps. Pixel below $2\sigma$ in the total intensity ($\sigma=11$~\mjybeam) or in the polarized intensity map ($\sigma=20$~\mjybeam) were blanked.}\label{fig:radio-pol}
\end{figure}

A map at 1400~MHz has been extracted from the NVSS \citep{Condon1998}. The image rms is $\sim0.5$~\mjybeam at a resolution of 45\arcsec. This high-resolution image of \target{} shows the extremely narrow front of the expanding shell towards the east (see Fig.~\ref{fig:VLA1400}). Although the extension of the SNR is smaller than the largest detectable angular scale for the VLA in D-configuration, flux density measurements are likely to be underestimated due to the poor coverage of the central part of the $uv$-plane caused by the short observing time. NVSS polarized intensity in the direction of \target{} is too weak to provide useful information.

A map at 2417 MHz has been obtained from the Parkes 2.4 GHz Survey of the southern Galactic Plane \citep{Duncan1997} with an image noise of $\sim11$~\mjybeam{} and a resolution of 10\arcmin (see Fig.~\ref{fig:PKS2400}). The target is detected but it is embedded in a larger-scale emission. Enhanced emission is visible in the corresponding regions of the shell that are detected in the higher-resolution maps. Compared to all other radio maps, the Parkes 2.4 GHz map shows entirely the extended structure of the source, having been extracted from a single-dish observation with no filtering applied. Polarized intensity maps from the Parkes 2.4 GHz survey have also been analysed. After applying a cut at $2\sigma$ both in the total intensity and in the polarized intensity maps, a region with a fractional polarization emission of $\sim 15\%$ has been detected in the north west part of the remnant (see Fig.~\ref{fig:radio-pol}).

We obtained a high-frequency radio map from the Parkes-MIT-NRAO \citep[PMN;][]{Griffith1993} survey at 4850 MHz with an image local noise of $\sim11$~\mjybeam{} and a resolution is of \am{4}{9} (see Fig.~\ref{fig:PNM4850}). The \target{} is visible even at this higher frequency but the elongated bow structure on the east is broken into two distinct parts. The presence of an extended emission inside the boundary of the SNR is also visible. The PMN survey was optimized for point source detection, consequently a low-pass filter was applied in the data reduction process \citep{Griffith1993} with the effect to bias the flux densities of detected sources, mostly if extended, towards lower values. We note therefore that also PMN flux densities are likely to be underestimated.


\begin{table*}[!t]
\caption{Radio observations}
\vspace{-15pt}
\label{tab:radio}
\begin{center}
\begin{tabular}{ccccccc}
Telescope & Frequency & Resolution & Largest scale & \multicolumn{2}{c}{Flux density} & Rms \\
          & (MHz)     &            & (arcmin)      & \multicolumn{2}{c}{(mJy)} & (mJy beam$^{-1}$) \\
          &           &            &               &  East & West & \bigstrut[b]\\
\hline
GMRT             & 325  & \beam{141}{89} & $\simeq 32$ & 151 & 280 & 1.5 \bigstrut[t]\\
VLA (NVSS)       & 1400 & \beam{45}{45}  & $\simeq 16$ & 48  & 67  & 0.5\\
Parkes 64m       & 2417 & \am{10}{23} $\times$ \am{10}{62} & -- & \multicolumn{2}{c}{$\sim3000^b$} & 11\\
Parkes 64m (PMN) & 4850 & \am{4}{9} $\times$ \am{4}{9} & $\simeq 57^{a}$ & 367 & 162 & 11 \bigstrut[b]\\
\end{tabular}             
\end{center}
$^a$ although the PMN is a single dish survey, a limit on the largest scale is given by the filtering procedure applied during data reduction \citep{Griffith1993}. $^b$ The target emission is blended with the surrounding Galactic extended emission.
\end{table*}

\subsection{DECam}
\label{sec:DECam}

\begin{figure}
\centering
\includegraphics[width=.9\columnwidth]{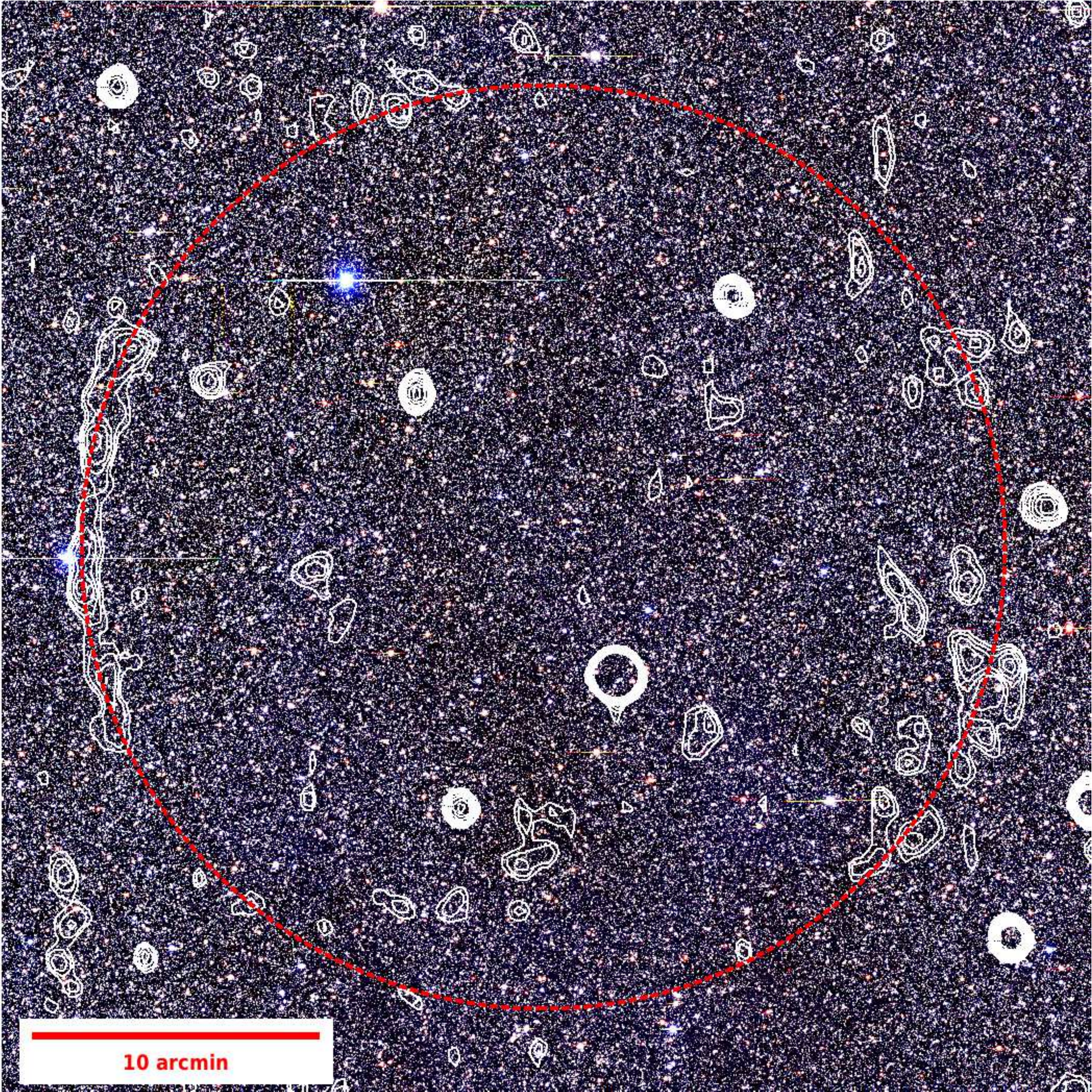}
\caption{DECam optical image (\textit{g}, \textit{r} and \textit{i} filters) of the target field. In red dashed-line the hypothetical size of the supernova remnant. Contours are from the NVSS survey (1400 MHz) at $\left( 2, 2.75, 3.5, 4.25, 5\right) \times \sigma$ with $\sigma = 0.5 $~\mjybeam.}\label{fig:opt1}
\end{figure}

In order to determine if the diffuse radio emission was a radio relic associated with a merging cluster of galaxies we observed the field in the optical during 4--7 Apr 2013 using DECam \citep{DePoy2008} on the CTIO 4 m Blanco telescope. We observed the field in $g$, $r$, and $i$ to facilitate star-galaxy separation and potentially select cluster members. Exposures in $g$, $r$, and $i$ were 100--200 s, dithered by 90\arcsec{} along each axis to cover the chip gaps. The total exposure times were 2000 s in $g$, 2000 s in $r$, and 2623 s in $i$.  These correspond to a $5\sigma$ point source AB magnitude depths of 26.6 in $g$, 26.1 in $r$, and 25.8 in $i$. After accounting for Galactic extinction \citep{Schlafly2011} the depths are 24.7 in $g$, 24.8 in $r$, and 24.8 in $i$. Using IRAF, all images were bias subtracted and flat-field corrected using nightly dome flats, then registered using \textit{msccmatch} for alignment. Residual gradients across the CCDs were removed using \textit{imsurfit}, and the resulting images were stacked using \textit{mscimatch} to estimate the relative normalizations. The final image is shown in Fig.~\ref{fig:opt1}. 

\subsection{Fermi-LAT}
\label{sec:fermi}

\begin{figure}
\centering
\includegraphics[width=\columnwidth]{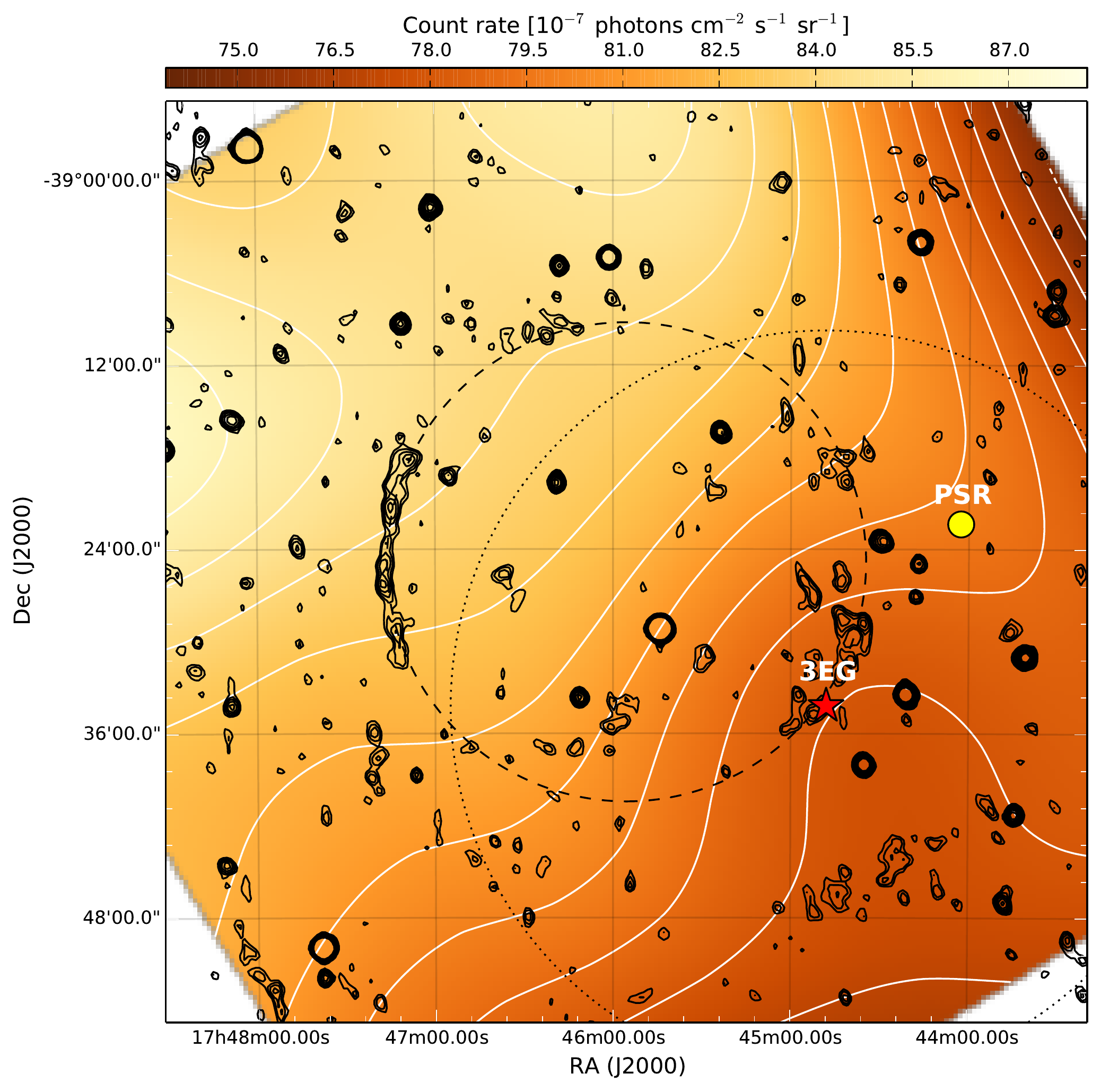}
\caption{Photon count from the Fermi-LAT observation in the energy band $1-500$ GeV. The angular resolution (68\% containment radius) increases monotonically from $\sim$1$^\circ$ to $\sim$0.2$^\circ$ in the energy range 1-300~GeV \citep{Ackermann2012}. White contours follow the $\gamma$-ray emission. Black contours follow the NVSS 1.4 GHz emission (Fig.~\ref{fig:VLA1400}). Markers and lines are as in Fig.~\ref{fig:VLA1400}.}\label{fig:fermi}
\end{figure}

To probe whether \target{} is a site of GCR acceleration, we searched for $\gamma$-ray emission at its location from the public database of the Fermi-LAT. LAT on board the Fermi Gamma-ray Space Telescope is able to detect $\gamma$-rays with energies between $100$~MeV and $300$~GeV \citep{Atwood2009}. Data used in this work were obtained between 2008 August 4 and 2013 November 3 (Pass 7 Reprocessed Weekly Files), which are available from the Fermi Science Support Center\footnote{\url{http://fermi.gsfc.nasa.gov/ssc/data/analysis/software/}}. We used the Fermi Science Tools ``v9r23p1'' package to reduce and analyse the data in the vicinity of \target. The region of interest was defined as a square $\pm3$\deg{} around the central point of \target. From the Fermi LAT data we extracted the $\gamma$-ray events for the region of interest in the energy range from 1 to 500 GeV. To calculate the flux we used the energy dependent point spread function of LAT \citep{Ackermann2012}. We did not considered energy events below 1~GeV due to the very poor angular resolution. The image with the photon count is shown in Fig.~\ref{fig:fermi}.

\section{\target}
\label{sec:target}

The candidate SNR has a radius of \am{15}{5} with a centre located at $\alpha=17$\hour45\min55\sec, $\delta=-39$\deg24\min50\sec\ (J2000) or $l=351.06$\deg, $b=-5.49$\deg (Galactic).

In Fig.~\ref{fig:GMRT325} and \ref{fig:radio} we present the four available radio images (325, 1400, 2417 and 4850 MHz) of \target. In the high-resolution NVSS image the very narrow edge of the expanding shell is visible on the east side. On the west side a more disrupted edge is visible. Hints of emission come also from the north-east side of the shell and are mostly visible in the GMRT and PMN maps. The radio emission is globally not symmetric and mostly located on the expanding shell, features typical of old SNRs where the interaction with the ISM has already modified the symmetry present in the primary expansion stages \citep[see e.g.][for IC443 and the Cygnus Loop respectively]{Green1986,Leahy1998}.

Given the radio morphology, another interpretation could be that the radio emission is a ``double radio relic'' generated by merging galaxy clusters producing shocks \citep{Feretti2012}. Relics are typically found in massive and merging systems with a hot intra-cluster medium ($10^7-10^8$ K) glowing in the X-ray. However, given the typical size of radio relics ($\sim 1$ Mpc), the cluster should be at $z\sim0.05$ and should have been detected by the ROSAT all sky survey although it is not. Furthermore, we find no evidence for significant galaxy over density associated with the radio emission in our optical observation (see Sec.~\ref{sec:DECam}). Given the limiting magnitudes of the optical images and the fact that diffuse cluster radio emission has only been observed in rich massive clusters we can confidently rule out any clusters associated with the radio emission below a redshift of 0.2. Were the radio emission to be associated with a redshift greater then 0.2, the projected physical scale of the emission would be $>16$ Mpc. This is $>7\sigma$ larger than the distribution of observed radio relics \citep{Feretti2012}. Thus it is highly unlikely that the diffuse radio emission is a radio relic associated with a merging galaxy cluster.

The extraction of a spectral index value is complicated because of the object's extension which could lower its flux density because of missing short baselines in interferometric observations and filtering in the single-dish case. This is clearly the case for the flux density extracted from the NVSS 1400 MHz map, which in both the east and west side of the source is lower than the flux density at 325 and 4850 MHz implying a concave spectrum not compatible with synchrotron radiation. 

 
\section{Discussion}
\label{sec:discussion}



\begin{figure}
\centering
\includegraphics[width=\columnwidth]{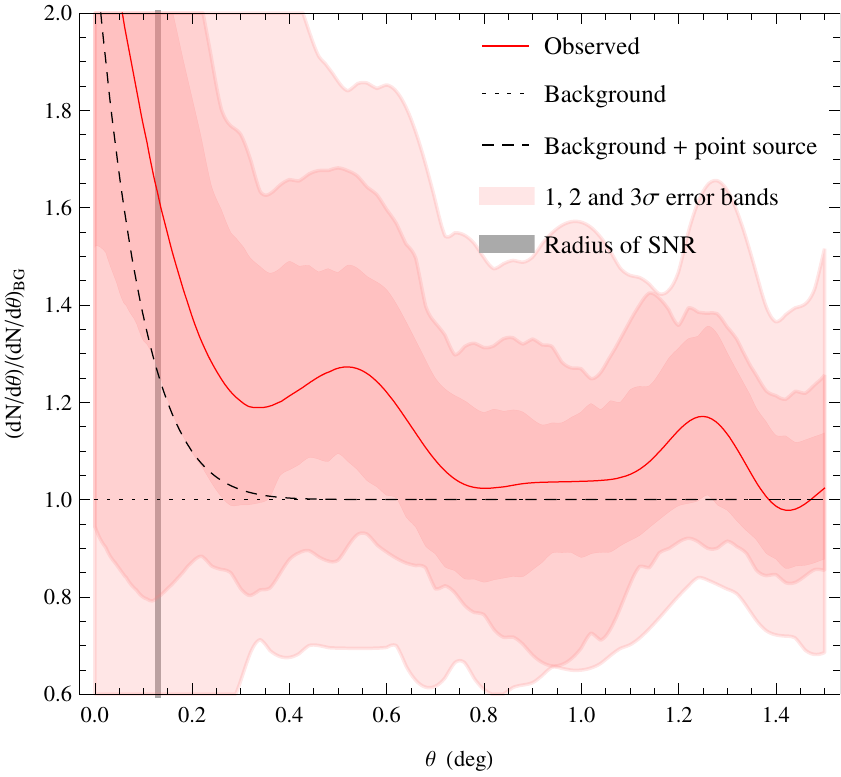}
\caption{Radial distribution of $\gamma$-ray events ($E > 10$~GeV) around the centre of the SNR. The red line denotes the ratio between the signal and the background estimated events (dotted line). Dashed line assumes a point source located at the centre of the SNR. A slightly spatially extended (confidence level $\sim$1.5$\sigma$) feature is related with SNR. Also, the two excesses can be seen at 0.5$^\circ$ and 1.2$^\circ$ from SNR probably related to the two Fermi LAT point sources at those distances \citep{Nolan2012}.}\label{fig:gamma_rad}
\end{figure}

The radio emission of \target{} is coincident with an unknown EGRET $\gamma$-ray detection (3EG J1744-3934) which could be the consequence of the interaction between the expanding shell and the surrounding medium, e.g. molecular clouds. The $\gamma$-ray source 3EG J1744-3934 \citep{Hartman1999} position, although having large errors ($r=0.66$\deg{} is the 95\% confidence area), is centred on the west side of \target. The source is marked as a possibly extended/multiple source and has a flux of $F(>100~\rm{MeV})=17.1\pm3.5 \times 10^{-8}$ photon cm$^{-2}$~s$^{-1}$ and a spectral index $\gamma = 2.42$. A new map of the $\gamma$-ray emission detected by Fermi-LAT in the energy range $1-500$ GeV in the area surrounding the nominal position of \target{} is reported in Fig.~\ref{fig:fermi}. We found no statistically significant $\gamma$-ray excess over the local background associated with \target{} in our study. The background is modelled as the averaged flux of region of interest, where the region ($<1^\circ$) of the SNR and the Fermi LAT point sources \citep{Nolan2012} are excluded. In the region associated with the SNR, the limiting photon flux at energies $> 1$ GeV is constrained to be $F(>1~{\rm GeV}) < 7.9 \times 10^{-9}$~cm$^{-2}$~s$^{-1}$. However, in the radial distribution of the $\gamma$-ray events around the centre of the SNR, a slightly spatially extended feature is detected with a confidence level of $\sim1.5 \sigma$, see Fig.~\ref{fig:gamma_rad}.

CO and \Hi{} maps obtained by NANTEN telescope do not show any association between the SNR and possible molecular cloud systems (Fukui Y. priv. comm.). If \target{} is indeed an old SNR, the $\gamma$-ray emission is absent because all CRs have escaped from the source, therefore energetic photons from neutral pion decay cannot be detected. This is in line with an absence of clouds along their path. If there were high density targets in the source surroundings, we could have seen GeV emission due to low-energy CRs (in the case of a target embedded in the source) or to high-energy CRs (in the case of a distant target) as in the case of the SNR W28 \citep{Giuliani2010}.

The pulsar PSR J1744-3922 is at a distance of 3.1 kpc, it was identified as a possible origin of the $\gamma$-ray emission for the source 3EG J1744-3934 \citep{Hessels2005} and it is close to the target SNR ($\sim23$\arcmin{} from the SNR centre towards west). A subsequent follow up with XMM showed no X-ray emission from the object and the source does not appear to be energetic enough to power the production of $\gamma$-rays \citep{Faulkner2004}.

The optical observation made with the DECam does not show any evidence of optical emission in any of the observed bands (see Fig.~\ref{fig:opt1}). The non-detection in the optical and in the $\gamma$-ray bands, together with the absence of CO and \Hi{} emission in the surroundings, the radio morphology and the low surface brightness, support the idea that \target{} must be an old SNR.


\section{Conclusions}
\label{sec:conclusions}

We report the discovery of an extended radio emission located in the proximity to the Galactic plane (Galactic latitude $b=-5.5$). The emission is visible in three radio surveys: NVSS at 1400 MHz, Parkes 64m at 2417 MHz, and PMN at 4850 MHz. The source has been also detected by the GMRT at 325 MHz. The source has not been detected in the optical band. Given the source morphology, we classify it as the old shell of a Galactic supernova remnant which no longer emits in the optical and designate it \target.

The presence of a SNR at these coordinates could also explain the unclassified EGRET $\gamma$-ray source 3EG J1744-3934 that may be the consequence of the interaction between particles escaping the remnant and the surrounding intra-galactic medium. However, deep analysis of the $\gamma$-ray data from Fermi-LAT did not show any significant excess in this region, no detection in CO or \Hi{} is also reported. In the radial distribution of the $\gamma$-ray events around the centre of the SNR, a slightly spatially extended excess is detected with a confidence level of $\sim1.5 \sigma$.


\begin{acknowledgements}
We would like to thank Ishwara Chandra C. H. and the staff of the GMRT that made these observations possible.
Based on observations obtained at Cerro Tololo Inter-American Observatory, a division of the National Optical Astronomy Observatories, which is operated by the Association of Universities for Research in Astronomy, Inc. under cooperative agreement with the National Science Foundation.
AH acknowledges the European Social Fund for the grants MTT8, MTT60 and the European Regional Development Fund for the grant TK120.
\end{acknowledgements}


\bibliographystyle{aa}
\bibliography{coccus}






\end{document}